# Simultaneous submicrometric 3D imaging of the micro-vascular network and the neuronal system in a mouse spinal cord

Michela Fratini[1,2], Inna Bukreeva[3], Gaetano Campi[4], Francesco Brun[5], Giuliana Tromba[6], Peter Modregger[7], Domenico Bucci[8], Giuseppe Battaglia[8], Raffaele Spadon[9], Maddalena Mastrogiacomo[9], Herwig Requardt[10], Federico Giove[1,11], Alberto Bravin[10] & Alessia Cedola[3*]

1. "Enrico Fermi" Centre MARBILab c/o Fondazione Santa Lucia Via Ardeatina, 306 00179 Roma, Italy

2. Department of Physic, University of Roma TRE, via della Vasca Navale, Rome, Italy

3. Institute for Physical and Chemical Process-CNR c/o Physics Department at 'Sapienza' University, Piazzale Aldo Moro 2, 00185 Rome, Italy

4. Institute of Crystallography-CNR, Monterotondo, Rome, Italy

5. Department of Engineering and Architecture, University of Trieste, Via A. Valerio, 10, 34127 Trieste, Italy

6. Elettra – Sincrotrone Trieste, Area Science Park, SS14, km164,5 – Basovizza, Trieste, Italy

7. SLS-PSI, Villigen, Zurich, Switzerland

8. I.R.C.C.S. Neuromed, Località Camerelle, 86077 Pozzilli, Italy

9. Department of Experimental Medicine, University of Genova & AUO San Martino - IST Istituto Nazionale per la Ricerca sul Cancro, Largo R. Benzi 10, 16132 Genova, Italy

10. European Synchrotron Radiation Facility, 71 Avenue des Martyrs, 38043 Grenoble, Cedex France

11. Department of Physics, Sapienza University of Rome, Piazzale Aldo Moro, 2, 00185 Roma - Italy

[*] Corresponding authors:
Alessia Cedola: Istituto per i Processi Chimico Fisici- CNR c/o Dipartimento di Fisica-Università Sapienza, P.le Aldo Moro 5, 00185 Roma, Italy
Tel.: +390649913991; Fax: +390649957617; E-mail address: alessia.cedola@.cnr.it

**Abstract**

Defaults in vascular (VN) and neuronal networks of spinal cord are responsible for serious neurodegenerative pathologies. Because of inadequate investigation tools, the lacking knowledge of the complete fine structure of VN and neuronal systems is a crucial problem. Conventional 2D imaging yields incomplete spatial coverage leading to possible data misinterpretation, whereas standard 3D computed tomography imaging achieves insufficient resolution and contrast.

We show that X-ray high-resolution phase-contrast tomography allows the simultaneous visualization of three-dimensional VN and neuronal systems of mouse spinal cord at scales spanning from millimeters to hundreds of nanometers, with neither contrast agent nor a destructive sample-preparation. We image both the 3D distribution of micro-capillary network and the micrometric nerve fibers, axon-bundles and neuron soma. Our approach is a crucial tool for pre-clinical investigation of neurodegenerative pathologies and spinal-cord-injuries. In particular, it should be an optimal tool to resolve the entangled relationship between VN and neuronal system.

The complex blood supply of the brain and spinal cord is particularly significant and relevant to both physiology and pathology of the central nervous system (CNS).

In fact, recent studies have shown that neurons, astrocytes, glia and microvessels seem to constitute a functional unit, the primary purpose of which is to maintain the homeostasis of the brain's microenvironment. Alterations of these vascular regulatory mechanisms lead to brain dysfunction and disease[1], including neurodegenerative pathologies, cancer and ischemia[2,3].

In particular, the neurodegenerative diseases, such as Alzheimer's disease, Parkinson's disease, amyotrophic lateral sclerosis, and Huntington's disease affect almost 300 Million of humans in the world and effective therapeutic treatments do not exist. One weak point of the research progress has been associated with the lacking knowledge of the structural relationship between the vascular network (VN) and the neuronal system. Because of the inadequate investigation tools, the physiology of the interactions between the neuroglial unit and the VN (including the mechanisms behind the functional hyperaemia) still needs to be fully elucidated. High resolution 3D imaging of the VN and cellular populations in a large volume of tissue, with a resolution sufficient to access the smallest capillaries and the neuron morphology, can bring to a better understanding of the cellular control of blood flow, and thus of the neuro-vascular coupling.

Up to date a number of different approaches are used to understand this issue, but all of them have serious limitations. Magnetic Resonance Imaging (MRI) or Positron Emission Tomography (PET) allow for structural and functional imaging. In particular these techniques allow the detection of alteration at biochemical andmesostructural level, which usually come before the anatomical alteration detected by other techniques. PET and fMRI provide quantitative maps of the changes in blood flow in the different anatomic structures. However, their spatial resolution limits raise the question of a possible micro-vascular organization at a scale smaller than the resolved one.

On the other hand, micrometric-scale imaging of brain tissue relies on 2D imaging techniques such as histological sectioning or Scanning Electron Microscopy (SEM). Well-consolidated and widely used techniques, they both provide impressive 2D images of the neuronal system or vessels. However, an invasive sample preparation is required in both cases for thinning the sample down to hundreds of microns. Although very informative at micron scale, these approaches yield incomplete spatial coverage and thus possibly lead to data misinterpretation. A major reason is that, albeit accurate, these methods are restricted to the investigation of a tissue volume within a finite depth depending on the incident radiation penetration ability. Moreover, imaging of large volumes of tissue in any microscope requires accurate sectioning of brain samples to an appropriate thickness, a procedure inherently destructive, and prone to induce damage within the imaged sections.

Post-mortem methods, such as the corrosion-cast technique, provide impressive pictures of the vasculature of the cortex and spinal cord using SEM, but do not offer the possibility of quantitative analyses[4] This is achieved only by complementing the SEM measurement with 3D micro-tomography[5] but this combination of techniques makes this approach of difficult application in pre-clinical investigations and moreover it requires an invasive sample preparation which can alter the 3D morphology of the system.

More recent techniques, such as two-photon microscopy, provide important *in vivo* 3D information and offer optical biopsy, for which high speed imaging is required. However, it is restricted by the low penetration depth and moderate size of the region of interest.

MR angiography and volumetric-computed tomography are powerful 3D techniques used to image arteries (and, less commonly also veins) in order to evaluate abnormalities. They have been applied to demonstrate the diffuse organization of tumor vessel architecture *in vivo*[6]. However, small vessels (smaller than 20 μm in diameter) cannot be visualized with either modality. Conventional X-ray angiography is commonly used to image neural vasculature in clinical practice, but it also has a detection limit, which is inappropriate for imaging the micrometric vessels and capillaries.

The high-resolution X-ray synchrotron phase contrast tomography (XSPCT) applied in combination with a contrast agent has been proposed for *ex-vivo* 3D imaging of micro-vascular networks inside several cubic millimeters of cerebral tissue and adapted to the analyses of primate vessel networks, as well as to brain tumors in an implanted rat model[7,8]. Such a method not only permits the imaging of volume portion of tissue, but it also allows a systematic digitalization of the vessels. This technique reaches spatial resolution close to microns, when invasive contrast agent, administered by *in vivo* perfusion methods, are used. Without contrast agent the detection limit for vessels has been shown being around 50 microns[9] Up to date, XSPCT appears the most promising *ex vivo* technique for a 3D imaging of the vasculature[10,11]. However, it is not yet applied without any contrast agent and/or with a sufficient resolution to access both the smallest capillaries and the neuron morphology.

The study of the interaction between VN and neuronal system is not be able to leave the capillary system out of consideration, being this the vehicle between the cells and the vessels.

Therefore, a 3D imaging spanning from millimeters to hundreds of nanometers, able to access a resolution to discriminate the smallest capillaries and the neuron morphology without invasive contrast agent and without aggressive sample preparation affecting possible data misinterpretation is still expected.

The results reported in this study are unique in providing a detailed analysis of the three-dimensional of the micro-vascular network and simultaneously the relevant interactions with neural cells of the healthy mouse's spinal cord, without any contrast agent. These results were obtained thanks to the applications of XSPCT exploiting the spatial resolution of 0.6 microns, the coherence of the X-ray source and the high image contrast coming from the phase technique. The data analysis and the 3D rendering enable the distinct visualization of the VN and neuronal systems in the same image.

We propose this experimental approach for the pre-clinical investigation of neurodegenerative pathologies and spinal cord injuries as well as for the study of physiological interactions within the neurovascular unit. Moreover, it is promising for solving the relationship between VN and neuronal system.

**RESULTS**

We show the simultaneous three-dimensional imaging of micro-vascular and neuronal system of the mouse spinal cord. Fig. 1a shows a 3D reconstruction of a 200 μm thick slab of the lumbar-sacral region of the spinal cord acquired with XSPCT without using contrast agent, with spatial resolution of 0.6 micron, obtained at TOMCAT beamline at SLS. The image shows an amazing contrast between *white and grey matter* of the spinal cord (inverted grey-levels in the figure): The "H" typical shape of the grey matter (of anterior horn and of the posterior horn)[12] can be well distinguished from the surrounding white matter. The gray matter bordering the central canal is well imaged The central and the radial vessels penetrating the gray matter are perfectly discernible and well imaged.

At the gray/white matter interface the white spots in the gray matter are the neuronal cell bodies (as discussed below) and thenerve fibers surround the gray matter, as confirmed by Nissl staining (Fig. 1b) and by immunohistological analysis of SMI-32, a marker of motor neurons (Fig. 1c).

XSPCT is able to provide a detailed image of the central radial vessels (from 10 to 15 microns of diameter) entering the gray matter from the outer part of the spinal cord (Fig. 1d). Immunohistochemistry images of laminin, a large extracellular matrix glycoprotein found in basement membranes of epithelia, surrounding blood vessels, show the presence of blood vessels in the anterior portion of the lumbar-sacral spinal cord obtained at different levels (Fig. 1e), very likely corresponding to the structure observed to XSPCT (Fig. 1d).

The unprecedented spatial resolution combined with the large field of view of the stack of tomographic slices allows to access information coming from both the VN and the neuronal system in the same image. In order to better investigate the VN of the spinal cord we performed an image segmentation of the Fig. 1a. The capillary (from 9 to $2.5 \pm 1$ μm in diameter) together with the intricate neuron fibers (from 5.5 to $9.5 \pm 1$ μm in diameter) are shown in Fig. 2a. Fig. 2b is a magnified region of Fig. 2a where the capillaries represented in red and the nervous fibers in green look differently: the vessels appear ramified while the nerve fibers bear a resemblance to "branches of a weeping Willow". Immunohistochemical analysis of spinal cord uses an antibody against myelin basic protein (MBP), a major constituent of the myelin sheath of oligodendrocytes and Schwann cells in the CNS and the peripheral nervous system. It shows the presence of neuronal fibers in the spinal cord, as shown in Fig. 2c, which parallel the complexity of neuron fibers imaged by XSPCT. Comparative images of spinal cord labelled with antibodies against laminin shows the vascular system (Fig. 2d). The images in Fig. 2c and 2d confirm the segmentation of Fig. 2a and 2b.

We compare the VN imaged without contrast agent with the case of a twin sample prepared with MICROFIL®, a compound that fills and enhances the opacity of the microvascular network. These images were recorded at the ID17 beamline at ESRF using a 3.5 µm pixel size detector. The architecture of the vascular network is shown in Fig. 3a (axial view) and 3b (longitudinal view). The spatial distribution of the VN in Fig. 3 confirms that there is a small peripheral and large central vascular supply that is characteristic of the lumbar-sacral region.

The very high spatial resolution of 0.6 µm imaging achieved in the experiment without contrast agent, allows the identification of a striking contrast between *white and grey* matter with single cells and nerve fibers being well distinguishable at the white/grey matter interface in the anterior horn of the spinal cord (Fig. 4a). This spinal cord region is highly enriched with somatic motoneurons motor neurons with a characteristic stellate morphology. Histology by Nissl staining (Fig. 4b) and hematoxylin/eosin staining (Fig. 4c) confirmed the images obtained by XSPCT. Moreover, immunohistochemical analysis with an antibody against the nonphosphorylated epitopes in neurofilament B (SMI-32) labelled motor neurons (Fig. 4d), which show the same morphology of Fig. 4a. Higher magnification images of SMI-32 labelled cells (Fig. 4f) and hematoxylin/eosin staining (Fig. 4g) parallel the morphology of a single motor neuron shown in Fig. 4e. The different grey levels of XSPCT image are proportional to different density inside the soma.

In Fig. 5a one nerve fiber is imaged at the interface with the grey matter, while Fig. 5b is a longitudinal projection of the sample at the same interface, where the connections between the cells are compatible with the dendrites branches.

Fig. 6b and 6c show the longitudinal cross section of the 3D reconstructed spinal cord in the region selected in Fig. 6a. In Fig. 6b three regions can be distinguished: region 1 is the central canal where the cerebral spinal fluid flows, region 2 is the grey matter, flooded by vessels of about 11 µm of diameter, and region 3 is the white matter. In the region 3 the axons running parallel to the spinal cord axis are rendered in white (better visualized in Fig. 6c). Typical diameter of a single nerve fiber within the CNS is > 1 micron.

## DISCUSSION

The possibility to investigate simultaneously the structure of the CNS and vessels network (Fig. 1), at a range of scale spanning from millimeters to sub-micrometer (Fig. 2,3) has a strong impact in a large number of pre-clinical investigations of pathologies, including neurodegenerative diseases and the regenerative medicine.

*Disorders of CNS* white matter are relatively common and exist in a myriad of forms[13] including multiple sclerosis[14] adrenoleukodystrophy (ALD), Pelizaeus–Merzbacher disease, and some metabolic disorders (e.g., Canavan disease, Krabbe disease)[15]. In particular neurodegenerative diseases have been strongly associated with vascular alterations[16,17,18]. Pathology studies have found that a higher prevalence of vascular disease in the brain may result in an increased premorbid diagnosis of AD[19] Therefore the study of the organization of the μVN and neuronal network is a crucial physiological issue, because neurovascular alterations are present in various neurodegenerative diseases.

On the other hand, *traumatic spinal cord injuries* induce microvascular changes that may contribute to secondary injuries and deficits observed in patients[20,21]. In particular, the resultant ischemia and the extravasation of the blood components contribute to a series of effects as edema formation, neuronal cell death, and damage to white matter tracts. Regenerative medicine[22], shows promise for the treatment of, traumatic diseases of the spinal cord.

In this framework it is evident the remarkable importance of the ex-vivo pre-investigations on animal models, which offers significant advantages to accelerate the process of drug/treatment discovery.

In this framework, a 3D imaging in an *ex-vivo* model with one single high resolution phase contrast tomographic measurement, displaying simultaneously the complete architecture of the VN, neuronal populations (Fig. 1), axon bundles up to a single neuron soma (Fig. 4 and Fig. 5), without neither contrast agent nor specific sample preparation, is of great interest.

The *disorders of CNS* and the *traumatic spinal cord injuries* are the most significant examples of applications which could take advantage of our approach, which is a crucial complementary tool for pre-clinical investigation and it would be able to solve the entangled relationship between VN and neuronal system.

## METHODS

*Animals*

Adult male C57Black mice (20-22 g, body weight) were purchased from Charles River, Calco, Italy and kept under controlled conditions (temperature: 22°C; humidity: 40%) on a 12-h light/dark cycle with food and water *ad libitum*. All experimental procedures were carried out according to the institutional ethical guidelines for the care and use of laboratory animals. Mice were divided into three groups: one group was perfused with MICROFIL®, a low-viscosity radio opaque polymer (Flow Tech, Inc., Carver, MA), which enhances the vascularization, one group was perfused with saline solution and one group was used for histology and immunohistochemistry analysis.

<u>Perfusion with saline solution and with MICROFIL®</u>

Mice were anesthetized by an intraperitoneal injection of ketamine (80 mg/kg)/xylazine (10 mg/kg) mixture and perfused transcardially with saline solution containing heparin (50 U/ml). One group was afterward perfused transcardially by MICROFIL® agent. At the end of perfusion, the animals were stored overnight at 4 °C to permit the solidification of the MICROFIL®. Afterwards spinal cords were dissected out, fixed in 4% paraformaldehyde for 24 h, and then maintained in 70% alcohol until analysis.

<u>Histology and immunohistochemistry</u>

For morphological evaluation, mice were sacrified and spinal cords were dissected out and immediately fixed for 24 h in ethyl alcohol (60%), acetic acid (10%) and chloroform (30%). After embedding in paraffin, spinal cord sections were cut at 20 μm and stained with thionin (Nissl staining) and hematoxylin/eosin (H&E).

For immunohistochemical analysis of MBP, mice were anesthetized by ketamine (100 mg/kg, i.p.)/xylazine (10 mg/kg, i.p.) and perfused transcardially with 4% paraformaldehyde in phosphate buffered saline. Spinal cords were removed, fixed for 24 h in ethyl alcohol (60%), acetic acid (10%) and chloroform (30%) and embedded in paraffin. Twenty μm sections were first soaked in 3% hydrogen peroxide to block endogenous peroxidase activity and then incubated overnight with rabbit polyclonal anti-MBP (1:2000, Chemicon International, Billerica, MA).

For immunohistochemical analysis of SMI-32 or laminin, mice were killed, spinal cords were removed, fixed for 24 h, as above, and embedded in paraffin. Twenty μm sections were first soaked in 3% hydrogen peroxide to block endogenous peroxidase activity and then incubated overnight with mouse monoclonal anti-SMI32 (1:1,000, Covance, Princeton, NJ) or rabbit polyclonal anti-laminin (1:200, Novus Biologicals, Littleton, CO). Slices were then incubated then for 1 h with secondary biotinylated anti-mouse or biotinylated anti-rabbit antibodies (1:200; Vector Laboratories, Burlingame, CA). 3,3-Diaminobenzidine tetrachloride was used for detection (ABC Elite kit; Vector Laboratories, Burlingame, CA). Control staining was performed without the primary antibodies.

*High Resolution Synchrotron Phase Contrast Tomography*

The image contrast in conventional X-ray imaging originates from the differential absorption properties of the sample. For weakly absorbing materials (like biological materials), the X-ray properties of tissues are too similar making it very difficult to differentiate them, in particular when high resolution (micrometric) imaging is applied.. In these cases, a better contrast can be achieved by imaging the phase modulation induced by the sample immersed in a coherent or partial coherent X-ray beam[23]. The application of the tomographic method provides the additional benefit to discriminate the different depths inside the sample and to provide a map of the different layers.

Different experimental approaches exist for detection of the X-ray phase contrast[24]. A simple yet effective phase contrast method is based upon in-line imaging after free space propagation. Nevertheless, the image captured by in-line propagation always contains mixed absorption and phase effects. Therefore, specific algorithm has been used to decouple absorption from phase information[25].

Phase contrast images were acquired with pixel size of 0.64 microns or 3.5 microns, respectively after administration or not of the MICROFIL® contrast agent. The phase retrieval algorithm proposed by Paganin et al.[25] was applied to all projections of the tomographic measurements.

*Experimental set-up*

The experiment of the un-stained sample was carried out at TOMCAT beamline at the Swiss Light Source (SLS) in Villigen (Switzerland). The monochromatic incident X-ray energy was 17 keV and a CCD camera with a pixel size of about 0.64 microns was set at a distance of 5 cm from the sample. The tomography has been acquired with 1601 projections covering a total angle range of 360°. The experiment of the stained sample was carried out at ID17 at ESRF in Grenoble. The monochromatic incident X-ray energy was 30 keV. The sample was set at a distance of 2.3 meters from the CCD camera with a pixel size of 3.5 µm . The tomography has been acquired with 2000 projections covering a total angle range of 360° with acquisition time of 1 second per point.

*Data analysis*

The phase retrieval algorithm[26] has been applied to the projections of the tomographic scans using the ANKAphase[27] code. The phase tomograms display different grey-level regions corresponding to different tissues.

The image segmentation to display the tissues independently and the reduction of the artifacts, have been performed using both commercial (Volview and ImageJ) and home-developed program software ( Matlab routines).

**Acknowledgments**

The authors are grateful to Laura Cancedda for helpful discussions during the paper preparation.

**Author contributions**

I.B. developed numerical code for data analysis; M.F., I.B., G.C., F.G., A.B. and A.C. followed the data analysis; M.F., G.C., G.T., F.B., P.M., H.R. and A.C. performed the experiments; R.S. and M.M. prepared the samples; D.B. and G.B performed the histological analysis and revised the paper;  M.F., A.B. and A.C.  wrote the paper.

**Additional information**

Competing financial interests: The authors declare no competing financial interests.

**FIGURE CAPTIONS**

**Figure 1**: a) X-ray Phase Contrast Tomography reconstructed volume of the lumbar-sacral region of the spinal cord. The image is obtained at a spatial resolution of 0.64 µm without contrast agent at TOMCAT beamline. b) Nissl staining of the lumbar-sacral spinal cord. c) Immunohistochemical analysis of SMI-32, a marker of motor neurons, in the lumbar-sacral region of the spinal cord. d) Detail of the radial vessels penetrating the gray matter. e) Immunohistochemistry of laminin, a marker of blood vessels, in the anterior portion of the lumbar-sacral spinal cord obtained at different levels. The red arrows in a) and b) indicate the central spinal cord canal.

**Figure 2**: VN investigation: a) X-ray Phase Contrast Tomography reconstructed volume of lumbar-sacral region of the spinal cord. The image, obtained with a pixel size of 0.64 µm without contrast agent at TOMOCAT beamline, was segmented to show the capillaries and the nerve fibers. b) Magnified region of a): the vessels are red and the nerve fibers are green. c) Immunohistochemical analysis of laminin, a marker of blood vessels, in the lumbar-sacral spinal cord showing a coronal section of the vascular system. d) Immunohistochemical analysis of myelin basic protein (MBP), a marker of the myelin sheath of nerve fibers in the spinal cord.

**Figure 3**: VN investigation: a) X-ray Phase Contrast Tomography reconstructed volume of lumbar-sacral region of the spinal cord with MICROFIL® as contrast agent. The image is obtained with a pixel size of 3.5 µm at ID17 at the ESRF. b) Longitudinal view of a).

**Figure 4**: Neural population investigation: a) The white/grey matter interface, imaged with inverted color, of a thick slab selected in the anterior horn of the lumbar-sacral spinal cord. b) Nissl staining, c) Hematoxylin/eosin staining and d) Immunohistochemical analysis of SMI-32, a marker of motor neurons, at the white/grey matter interface of the anterior horn of the lumbar-sacral spinal cord. e) Magnification of a single neuronal cell. Zoom of image f) SMI-32 labeled cells and g) hematoxylin/eosin staining showing a single neuronal cell.

**Figure 5:** a) One nerve fiber is imaged at the interface with the grey matter. b) Longitudinal view of the sample at the same interface.

**Figure 6**: a) X-ray Phase Contrast Tomography image of the lumbar-sacral region of the spinal cord, where the central region containing the empty canal is marked. b) Longitudinal slice of the sample in the region selected in a). The regions 1, 2, 3 highlight the central canal, grey matter and white matter, respectively. c) Zoom of the image at the interface between the grey matter and the longitudinal axons.

**Figure 1**

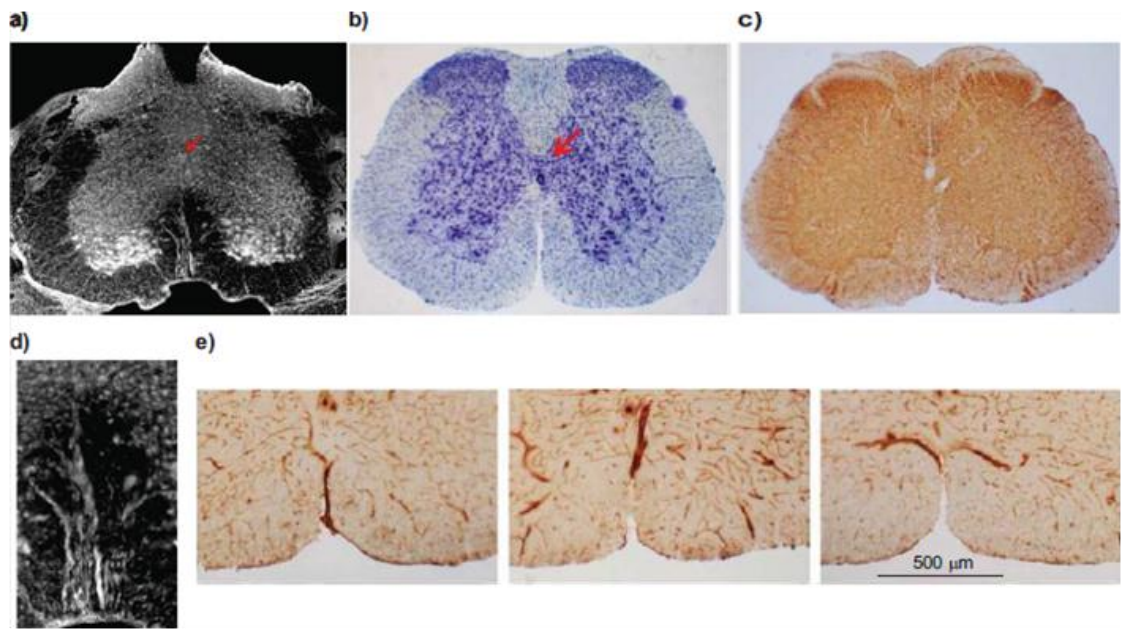

**Figure 2**

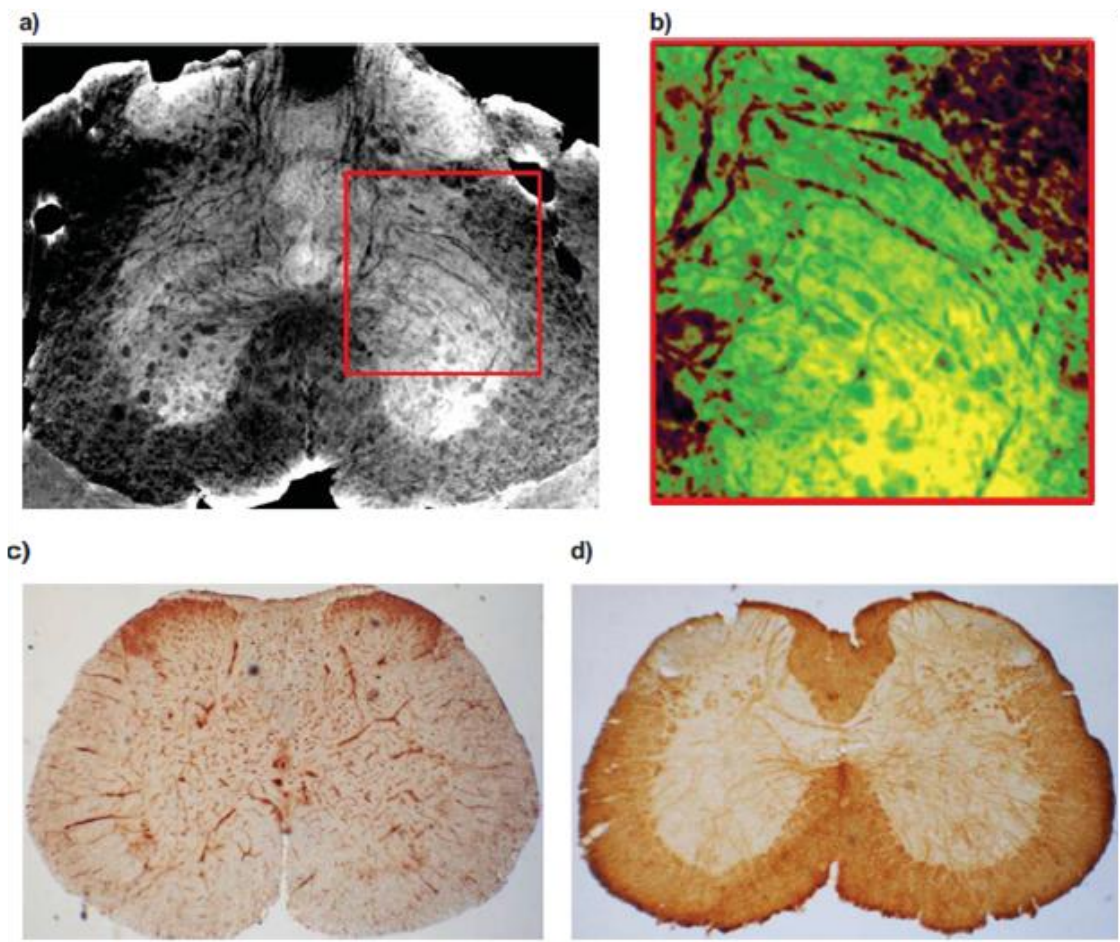

**Figure 3**

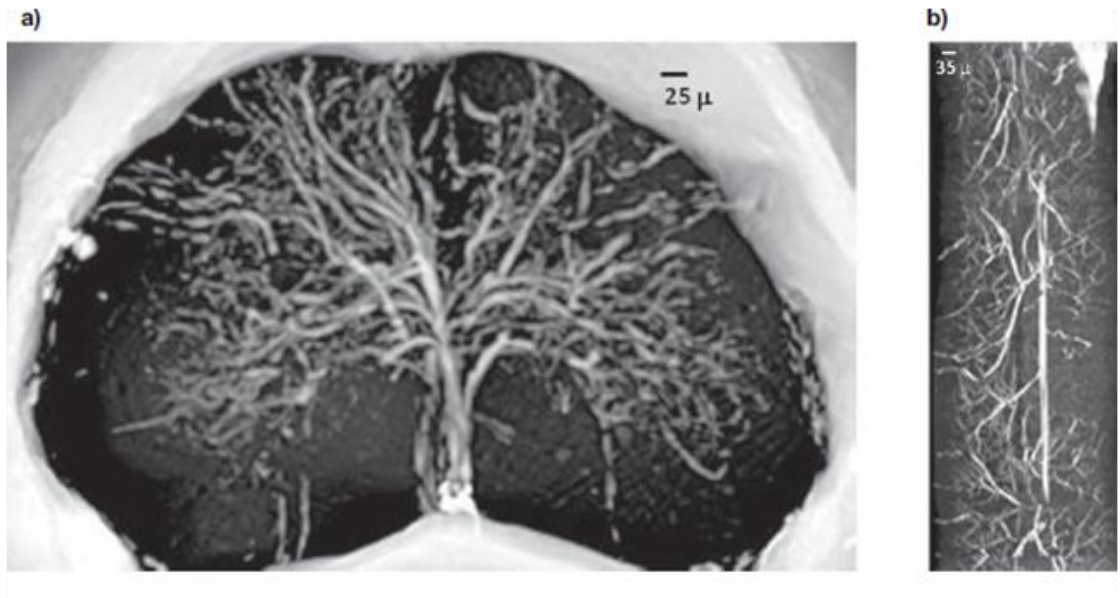

**Figure 4**

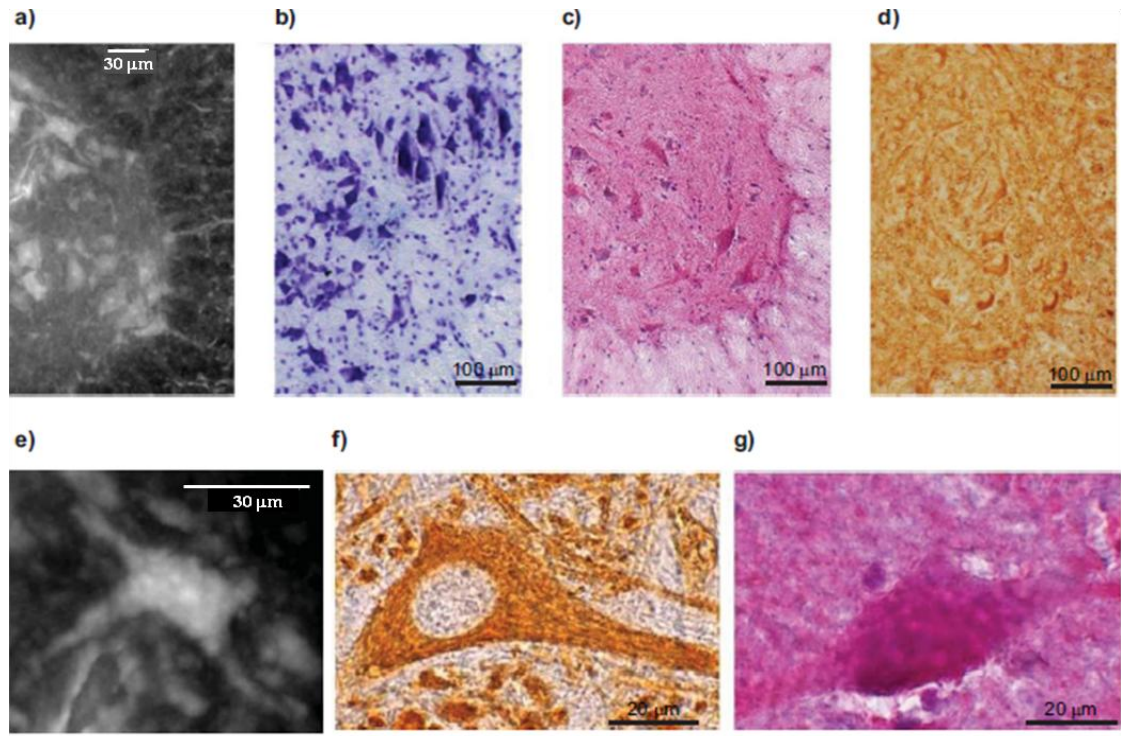

**Figure 5**

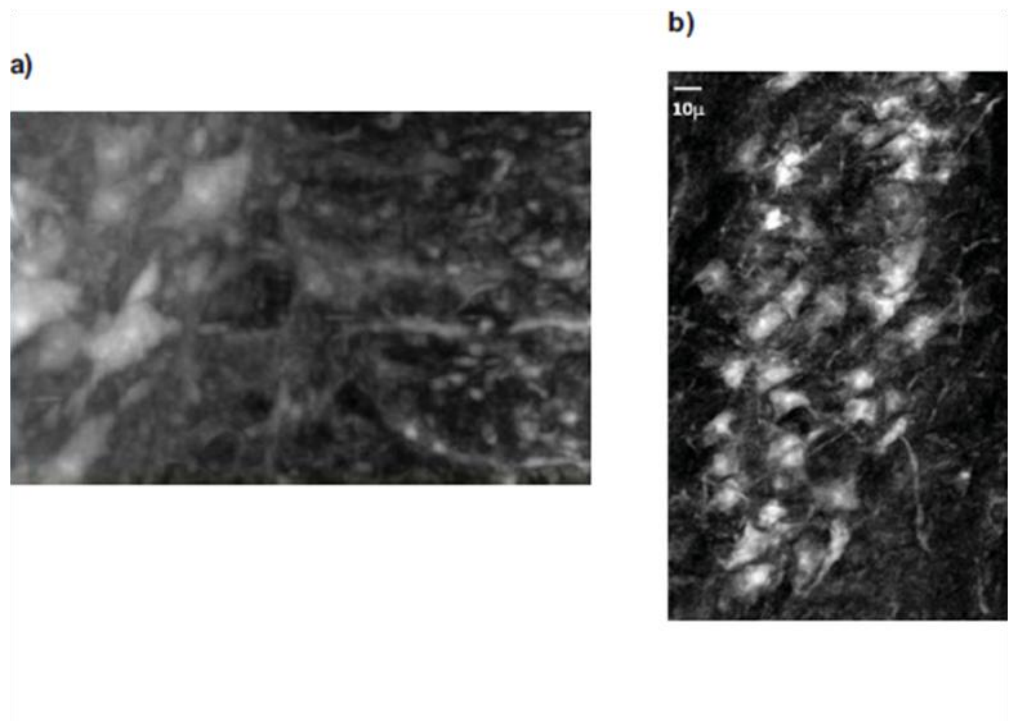

**Figure 6**

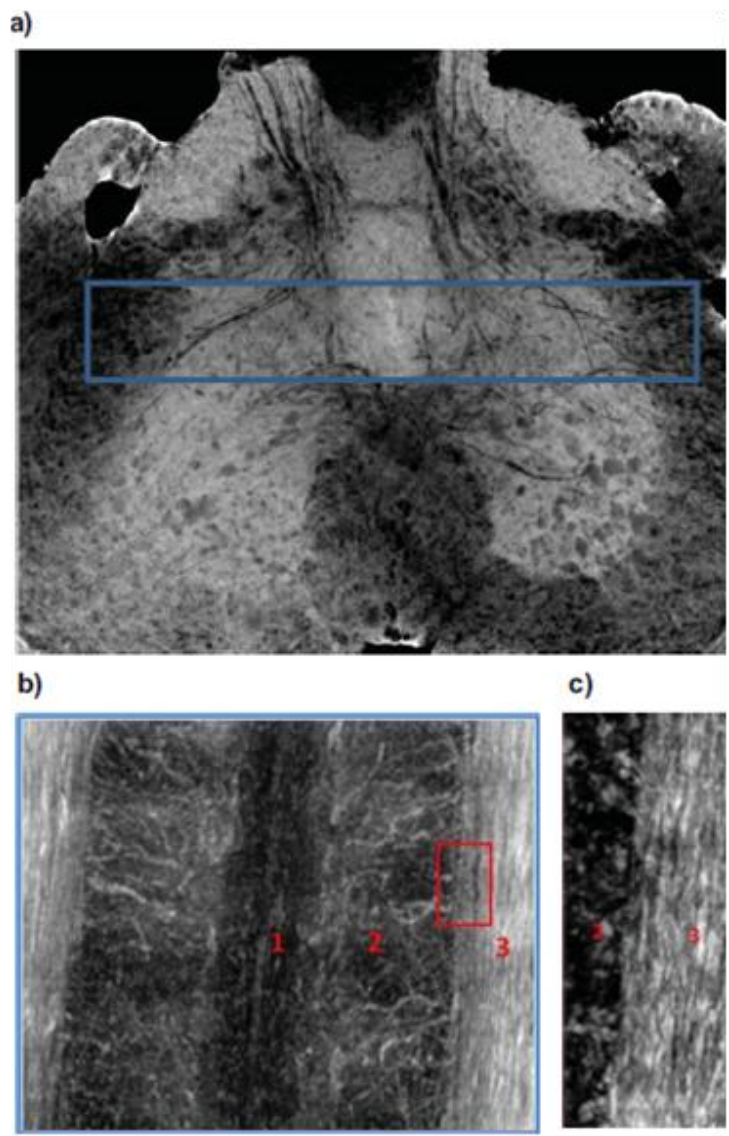